\documentclass[american,reprint,aps,pra,superscriptaddress,showpacs]{revtex4-1}
\usepackage{graphics}
\usepackage[T1]{fontenc}
\usepackage{amsmath,amsfonts,amsthm}
\usepackage[utf8]{inputenc}
\usepackage{newpxtext,newpxmath}
\usepackage[normalem]{ulem}
\usepackage{color}

\usepackage[normalem]{ulem}
\usepackage{amsmath}
\newcommand{\stkout}[1]{\ifmmode\text{\sout{\ensuremath{#1}}}\else\sout{#1}\fi}

\begin{document}
\def\parsedate #1:20#2#3#4#5#6#7#8\empty{20#2#3/#4#5/#6#7}
\def\moddate#1{\expandafter\parsedate\pdffilemoddate{#1}\empty}
\bibliographystyle{apsrev}

\title{ Decoherence and  Visibility Enhancement in Multi-Path Interference}

\author{Sandeep Mishra}
\email{sandeep.mtec@gmail.com}
\author{Anu Venugopalan}
\email{anu.venugopalan@gmail.com}
\affiliation{University School of Basic and Applied Sciences, G.G.S. Indraprastha University, Sector 16C Dwarka, Delhi - 110078, India.}
\author{Tabish Qureshi}
\email{tabish@ctp-jamia.res.in}
\affiliation{Centre for Theoretical Physics, Jamia Millia Islamia, New Delhi-110025, India.}

\begin{abstract}

We theoretically analyze the observations reported in  a four-path quantum
interference experiment  via multiple beam Ramsey interference
[Phys. Rev. Lett. 86, 559 (2001)]. In this experiment, a selective scattering of
photons from just one interfering path  causes decoherence. However, contrary to expectations, there is an increase in the contrast of the interference pattern, demonstrating that
path selective decoherence can not only lead to a
decrease,  but under certain conditions, to an {\em increase} of the fringe contrast. Here we
explain this seemingly counter-intuitive effect based on a model for a
multi-path interference, with four to six slits, in the  presence of
decoherence. The effect of the environment is modeled via a coupling to
a bath of harmonic oscillators. When decoherence is introduced in one of the multiple paths,
an  enhancement in fringe contrast is seen under certain conditions.
A similar effect is shown to appear if instead of path-selective decoherence, a
selective path detector is introduced.  Our analysis points to
the fact that while  traditional fringe visibility captures  the wave nature
in the two-path case, it can fail in multi-path situations.
We explain the enhancement of fringe visibility and also show that
{\em quantum coherence} based on the $l_1$ norm of coherence,
in contrast to traditional visibility, remains a good quantifier of wave
nature, even in such situations.  The enhancement of fringe contrast
in the presence of environmental decoherence underscores the limitations of traditional visibility as a good measure for wave nature in quantifying complementarity and also makes it an unlikely
candidate for quantifying decoherence. Our analysis  could lead to better insight in ways to quantify decoherence in multi-path interference, and in studies  that seek to exploit quantum superpositions and quantum coherence for quantum information applications.

\end{abstract}

\maketitle

\section{Introduction} \label{intro}
Quantum interference lies at the heart of quantum phenomena and is a
subject of continual and intense interest, fascination and study. Feynman
{\em et al.} placed the double-slit interference experiment at the
very core of quantum mechanics and famously declared that it contains
"the only mystery"  of quantum mechanics \cite{feynman}. While the
double-slit experiment for classical light is satisfactorily understood
in terms of the wave nature of light, interference experiments in
quantum systems continue to throw up conceptual challenges just
as advances in  technology push  frontiers to implement  hitherto
impossible interference experiments with larger and larger quantum
systems like cold atoms,  large molecules  and Bose Einstein condensates
\cite{neon,c60,virus,hackermuller,rev,horn,10000amu}. Larger and larger
interfering quantum objects are also vulnerable to decoherence which is
believed to wash out quantum behavior and drive  the transition
from  quantum to classical systems \cite{zeh,zurek,max}. Germane
to the debate surrounding the double-slit experiment in quantum
mechanics is Bohr’s {\em Principle of Complementarity} \cite{bohr},
and discussions regarding the consequences of ”which-way”  detection
on the interference pattern. The observation of an interference pattern is a signature of the wave
nature of the quanton while  the acquisition of which-way information
marks its particle nature, both of which are considered complementary or
mutually exclusive. In other words, the complete ignorance of  which
way information gives a fringe pattern with perfect visibility while
the presence of full which way information washes out the fringes. In
an intermediate regime where one obtains {\em incomplete} which way
information,  interference fringes persist with  {\em reduced}
visibility. This was established in a pioneering result by  Wootters and
Zurek \cite{w_zurek}, following which, Greenberger and Yasin \cite{GY}
and  Englert \cite{englert1, englert2} established  complementarity
relationships between measures which quantitatively estimate the
which path information, and the visibility, which measures the fringe
contrast. Subsequent studies involved the inclusion of which way detectors
in the interference experiment and the concept of the quantum eraser
\cite{qerasure,kim,walborn,kwait} and complementarity relations in
two-path interference were also successfully validated by experiments
\cite{chapman,buks,durr_t,bertet,jacques}. Central to all investigations
involving duality relations has been the traditional notion of visibility
as a measure of the wave nature given by
\begin{equation} \label{visibility}
{\mathcal V} = \frac{I_{max}-I_{min}}{I_{max}+I_{min}},
\end{equation}
where $I_{max}$ and $I_{min}$ are the  maximum and minimum of the
interference pattern (intensity distribution) on the screen. Following
the double-slit studies, it  was natural to explore  analogous forms
of interferometric duality when there is a multi-path interference
by the quanton. Some significant studies  in this direction were made
\cite{durr,bimonte1,asad,bera,coles,biswas,roy}, particularly
by D{\"u}rr \cite{durr} where the inequality relations
characterizing wave and particle properties in two-beam interferometers
were generalized to multibeam interferometers. For the particular case
of three-slit interference, Siddiqui and Qureshi have recently derived
a new duality relation \cite{asad}. However, inspite of these results,
it is being increasingly doubted whether the traditional notion of
visibility is  really compatible with the intuitive idea of complementarity, except  in the case of two-slit interference \cite{MW_1,MW_2,bimonte2,luis}.  Inspite of several studies,
duality relations involving interference fringe visibility and
path distinguishability which are believed to be complementary to each other,
have not been derived so far for an $n$-path quantum interferometer.
Our analysis of Mei and Weitz's experiment \cite{MW_1,MW_2} in this paper
will also highlight the inadequacy of visibility as a satisfactory
measure of wave nature, as understood conventionally.
 
The notion of coherence  initimately captures the wave aspect of
both classical light and quantum systems.  Recently, a measure for
quantum coherence was defined in the framework of quantum information
theory by Baumgratz {\em et al.} \cite{coherence}. This was used in defining
a normalized quantum coherence measure by Bera {\em et al.} \cite{bera} to
derive a  wave-particle duality relation for arbitrary multipath quantum
interference phenomena. Simply stated, the normalized coherence, also
known as the $l_1$ norm of coherence, is defined as the sum of the
absolute values of the off-diagonal elements of the density
matrix describing the quantum system:
\begin{equation}
{\mathcal C} = {1\over n-1}\sum_{i\neq j} |\langle i|\rho|j\rangle|,
\label{coherence}
\end{equation}
where $n$ is the dimensionality of the Hilbert space, and the value of
$\mathcal{C}$  always lies between 0 and 1. This measure is clearly
basis dependent and $\rho_{ij}=\langle i|\rho| j\rangle$ are the matrix
elements of the density operator of the system,
in the basis formed by the set of $n$ orthogonal states, which correspond
to the quanton passing through the $n$ different slits. Using the above
measure, Bera {\em et al.} \cite{bera} stated the Bohr complementarity principle
for multipath quantum interference  as a duality relation between {\em
quantum coherence} (in contrast to  conventional visibility) and  path
distinguishability, whose measure was described in a particular fashion
via unambiguous quantum state discrimination (UQSD) \cite{uqsd1,uqsd2}. The
result of Bera {\em et al.} firmly established that quantum coherence and
path distinguishability are truly complementary in nature, thus making
quantum coherence superior to the notion of traditional visibility
in a satisfactory understanding of wave particle duality. It is
worthwhile to point out here that the two-slit interference remains
an exceptional situation where the traditional visibility is the
same as quantum coherence. Following the work of Bera {\em et al.}, Paul
and Qureshi presented a method of experimentally measuring quantum
coherence in a multislit interference experiment \cite{tania}, putting
it on a firmer footing as a true experimentally measurable signature
of wave nature. Recently, Venugopalan {\em et al.} \cite{venu_2019} analyzed
the  multi-slit interference experiment with which-way detectors in
the presence of environmental coupling  and obtained  an expression for
quantum coherence as a function of the parameters of the environment,
thus providing a way to quantify decoherence through a measurement of
quantum coherence. While all these studies seem to undermine the role of
traditional visibility vis a vis quantum coherence, there is no denying
the fact that traditional visibility remains an intuitive and directly
observable and measurable sign of the underlying wave phenomenon in a wide
variety of experiments. Indeed, for a two-slit interference, a decrease
in the Michelson contrast  (traditional visibility) is a certain sign of
which-path information or the presence of decoherence. In view of this,
it does seem difficult to discard traditional visibility completely in
favour of coherence for situations involving more that two-paths.

In the following we  theoretically analyze the observations reported in
a four-path quantum interference experiment  via multiple beam Ramsey
interference \cite{MW_1, MW_2}. In this experiment, a selective scattering of photons
from just one interfering path  causes decoherence, revealing the which
path information, demonstrating that path selective decoherence can
not only lead to a decrease but, under certain conditions,  an {\em increase} of the fringe contrast. We analyze the experiment and explain this seemingly counterintuitive effect based
on a model for a multi-path interference  in
the  presence of decoherence. We show that  an enhancement of fringe
visibility  with path selective decoherence and $\pi$ phase in one path
can be seen where the results are particularly interesting for three and
four paths. However, we find that with a further increase in the number
of interfering paths, the effect is not that significant.  Our analysis
re-affirms the fact that while  traditional fringe visibility captures
the intuitive idea of complementarity in the two-path case (where an
increase in which path information  implies a decrease of fringe visibility for pure states), it can fail in multi path situations and under the  particular conditions of \cite{MW_1, MW_2},
it  can show unexpected and counterintuitive behaviour. We  correlate
the effect to  the recently introduced $l_1$ norm of {quantum coherence}
which can be experimentally measured and show that
coherence, as defined by  the $l_1$ norm, in contrast to traditional
visibility, captures complementarity and remains a good
quantifier of wave nature. Further, we show  that even in  situations
where visibility could increase with increasing decoherence (or which
path information), coherence always decreases and preserves its part in
the complementarity relations. Moreover, in such situations, the amount
of decoherence can be estimated only by a measure of the residual
coherence and cannot be inferred by changes (increase or decrease) in fringe contrast.

The rest of the paper is organized as follows: In Section \ref{multislit}
we describe a simple model that captures a multi-path interference
set up with which way detectors with the introduction of a $\pi$ phase
and the possibility of which path information in one of the paths, to
mimic the experiment of Mei and Weitz \cite{MW_1,MW_2}. We look at the
intensity distribution for  $n$ paths and  analyze the cases for three
and four paths. We explain the enhancement of visibility with which path
detection (and $\pi$ phase) in one path and compare it with the {$l_1$ norm of } quantum
coherence. In the three path case we show that there is initially a
decrease in visibility as path information increases followed by an
{\em increase}. We find that complete path information in one path does
not lead to an overall enhancement of visibility compared to the case
when all paths are indistinguishable. However, a {\em partial} path
information can lead to visibility enhancement as will be clear in the
following. In the four path case  we see that there
is initially a decrease in visibility as path information increases,
following which there is a significant {\em increase} in the visibility
with increasing path information in tune with the observations seen
in the four-path experiment of Mei and Weitz. Similar trends are seen
in the case of five and six slits. In Section \ref{multislit_dech}
we incorporate the effect of a real environment modelled by a bath of
harmonic oscillators. This environmental coupling mimics the effect of
photon scattering in a single path and we add a $\pi$ phase to create
the conditions of Mei and Weitz's experiment \cite{MW_1, MW_2}. Once again, our analysis
shows results predicted in Section \ref{intro}. We correlate our results
for visibility with quantum coherence and decoherence and  discuss its
implications in Section \ref{results} before concluding in  Section
\ref{conclusions}.

\section{Multislit interference with which way path detectors} \label{multislit}

\subsection{Visibility}

Let us consider a quanton passing through $n$ paths (slits) such that
the quantum state corresponding to the $n^{th}$ path is $\vert \psi _{n}
\rangle$. Then the state of the quanton can be written as
\begin{equation}
|\Psi\rangle = c_1|\psi_1\rangle + c_2|\psi_2\rangle  +  c_3|\psi_3\rangle \dots  +c_n|\psi_n \rangle,
\end{equation}
where {$\sum _{i} ^{n} \vert c_{i}\vert ^2= 1 $} and the states $\vert \psi _{n} \rangle$
are orthonormal to each other.  Let this quanton be entangled with an
ancilla system which can be considered as an effective environment or any
other path detecting device. The states of the ancilla $\vert \chi _{n}
\rangle$ are assumed to be normalized but are not necessarily orthogonal
to each other. The density operator corresponding to the combined state
of the quanton and the ancilla after their interaction is the entangled
state:
\begin{equation}
\rho = \sum _{j=1} ^{n} \sum _{k=1} ^{n} c_{j}c^{*}_{k} \vert \psi _{j} \rangle \langle \psi _{k} \vert e ^{i(\theta_{j}- \theta_{k})} \otimes \vert \chi _{j} \rangle \langle \chi _{k} \vert, 
\end{equation} 
where $\theta_{j}$ represents the phase by which the beam coming out
of the $j^{th}$ path is shifted. Since we are only interested in the
dynamics of the quanton, we will trace over the ancilla states to obtain
the density operator of the {\em reduced}  state of the quanton  as:
 \begin{equation}
\rho_{red} = \sum _{j=1} ^{n} \sum _{k=1} ^{n} c_{j}c^{*}_{k} \vert \psi _{j} \rangle \langle \psi _{k} \vert e ^{i(\theta_{j}- \theta_{k})} \otimes \langle \chi _{k} \vert \chi _{j} \rangle. 
\end{equation} 
Let us assume that after emerging from the $n$ slits, the beams are
combined and split into
new channels, whose states may be represented by $\vert \phi_{i} \rangle$.
For simplicity, we assume that all the original beams have
equal overlap with a particular output channel and that this overlap is
given by $\alpha$. The probability, $I$, of finding the quanton in the
$i^{th}$ channel is then
\begin{eqnarray} \label{I_nslit}
I &=& \langle \phi_{i} \vert \rho_{red} \vert \phi_{i} \rangle \\ \nonumber 
  &=& \vert \alpha \vert^{2} \left( 1 + \sum_{j\neq k} \vert c_{j} \vert \vert c_{j} \vert \langle \chi _{k} \vert \chi _{j} \rangle \cos(\theta_{j}-\theta_{k}) \right).
\end{eqnarray}

\begin{figure}
\centerline{\resizebox{9cm}{!}{\includegraphics{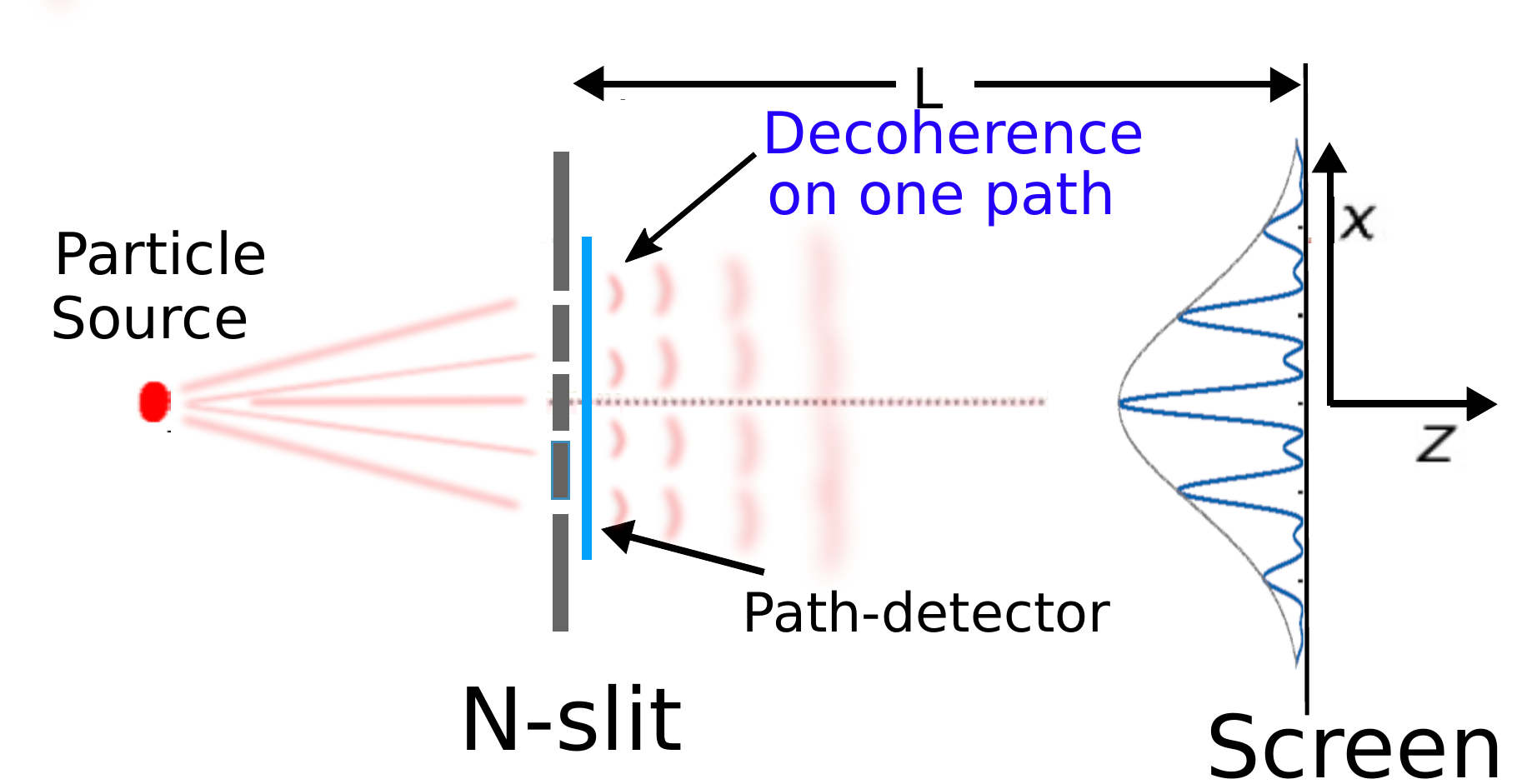}}}
\caption{Schematic diagram of a n-slit interference experiment, with 
a quantum path-detector. The interfering  quanton is affected by  
interaction with an environment only in one path  which has an additional phase of $\pi$.}
\label{nslit}
\end{figure}

Further, it is also assumed that the phases in the beams can be
independently varied. We now look to incorporate the conditions of
the experiment of Mei and Weitz \cite{MW_1, MW_2}, i.e., introduce the
possibility of {\em which path information} and a $\pi$ phase in only
one path. Let this be the $n^{th}$ path. This means that  the rest of the
$n-1$ detector states are identical with each other (i.e., their overlaps
$\langle \chi _{k} \vert \chi _{j}\rangle=1$) but these $n-1$ detector
states all have an overlap with the $n^{th}$ detector states given by $
\langle \chi _{j} \vert \chi _{n} \rangle = \beta$, where $j= 1,2,3 ...,
n-1$, and the value of $\beta$ lies between 0 and 1. We can represent
the path information of the $n^{th}$ path by a quantity {\em one-path
knowledge} which can be parametrized by $1-\beta$. If $\beta=0$, it
would imply complete one-path knowledge, with value 1. On the other hand,
if $\beta=1$, it implies zero one-path knowledge and all
paths are indistinguishable. To mimic the conditions of the experiment,
we add a $\pi$ phase to the $n^{th}$ detector. Further, for simplicity we
assume that the phases associated with the $n$ paths are such that they
can be written as $\theta_{k} = k \theta$ and $\theta_{n} = n\theta +
\pi$. The density operator for this  reduced state of the  quanton can then be
 represented as the  $n \times n$ matrix
\begin{equation} \label{state}
\rho_{red} =  \frac{1}{n}\left(
     \begin{array}{ccccccc}
      1 & 1  & 1 &  1 & .&.& -\beta \\
      1 & 1  & 1 &  1 & .&.& -\beta \\
      1 & 1  & 1 &  1 & .&.& -\beta \\
      1 & 1  & 1 &  1 & .&.& -\beta \\
      . & .  & . &  . & .&.& -\beta \\
      . & .  & . &  . & .&.& -\beta \\
      -\beta & -\beta  & -\beta &  -\beta & -\beta & -\beta & 1 \\
     \end{array}
   \right), 
\end{equation}
and the  intensity (\ref{I_nslit}) of the quanton in the $i^{th}$ channel can be shown to be:
\begin{eqnarray} \label{intensity}
I &=& \vert \alpha \vert^{2} \left[ 1 + \sum_{j = 3}^{n} (n+1-j)\cos \left( (j-2)\theta \right)  \right.\nonumber\\
&& \left. -\frac{2\beta}{n} \cos \left( (n-1) \theta \right) - \frac{2\beta}{n} \frac{\cos \left( \frac{(n-1) \theta}{2} \right) \sin \left( \frac{(n-2) \theta}{2} \right)}{\sin(\frac{\theta}{2})}\right].
\end{eqnarray}
One can see from (\ref{intensity}) that the intensity for the two-path
case ($n=2$ ) is  $\vert \alpha \vert^2 \left( 1- \beta \cos\theta \right)$. As expected,
the visibility, (\ref{visibility}),  ${\mathcal V} =  \beta $, for the
two path case. Clearly,  for full one-path knowledge ($\beta=0$), the
visibility is $0$ while for zero one-path knowledge ($\beta=1$) visibility
is $1$.  As mentioned before, the two-path case is straightforward and
in tune with all conventional notions of visibility and complementarity.
\begin{figure}
\centerline{\resizebox{9cm}{!}{\includegraphics{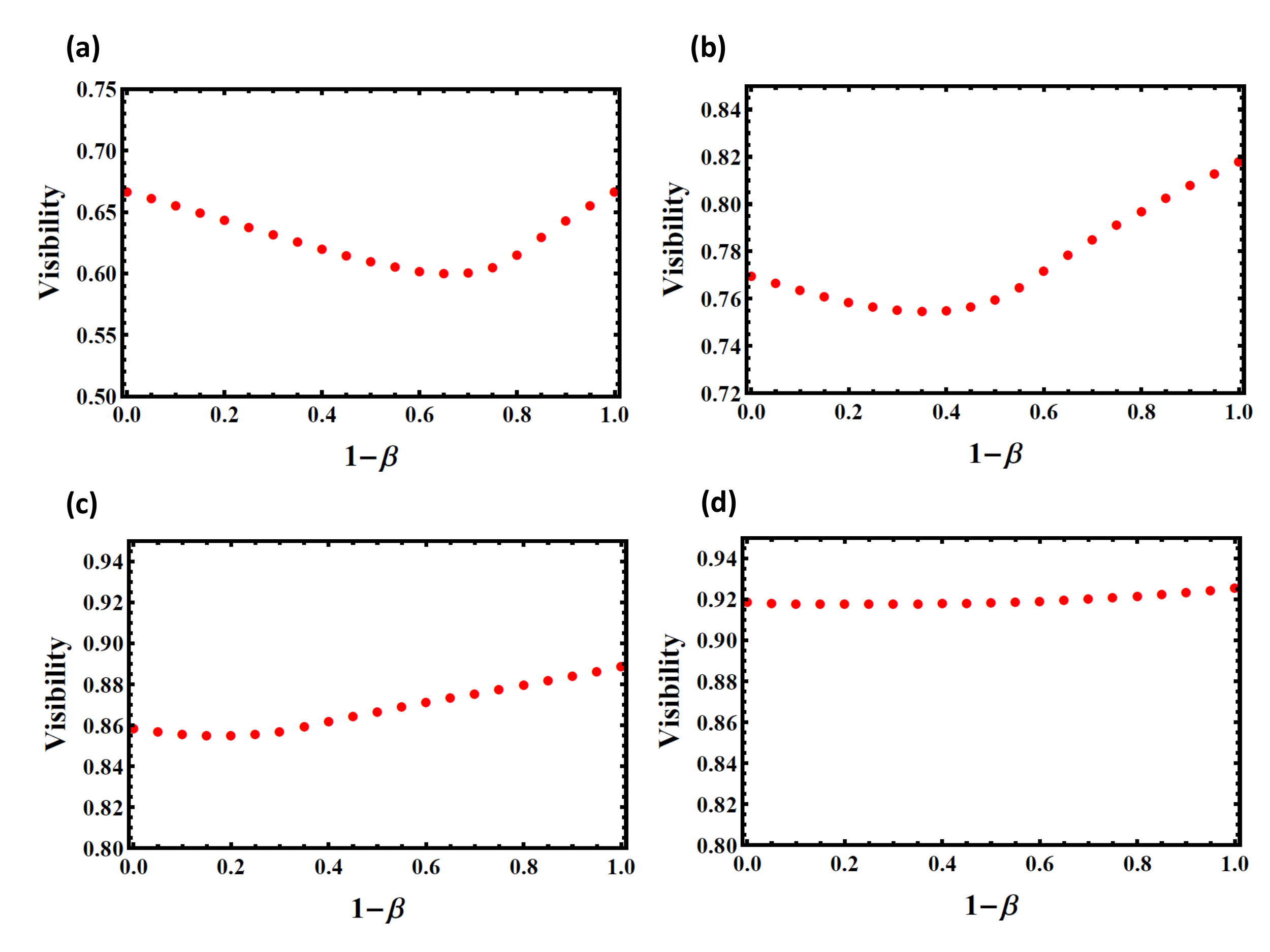}}}
\caption{Variation in visibility with respect to one-path knowledge ($1-\beta$)
for (a) 3 path (b) 4 path (c) 5 path  and (d) 6 path interference.
}
\label{visibility_fig}
\end{figure}
Let us now look at the scenario where there are more than two paths
($n\geq 3$). For $n=3$, one can see that  intensity (\ref{intensity})
is given by:
\begin{equation}\label{3_slit}
I= \vert \alpha \vert^{2} \left[ 1 + \frac{2(1-\beta)}{3} \cos(\theta) -\frac{2\beta}{3} \cos(2\theta)\right].
\end{equation}  
Note that (\ref{3_slit}) incorporates the possibility of path
information in the third path (via $\beta$), and this path also has
an additional $\pi$ phase. From (\ref{3_slit}), the visibility as a
function of the detector overlap $\beta$ (which quantifies the one-path knowledge) is shown in Fig. \ref{visibility_fig}(a). One can see  that
for both the case of complete path information ($\beta=0$) and no path
information ($\beta=1$), the visibility, (\ref{visibility}),  for the
$n=3$ case is  $\frac{2}{3}$. Thus, in contrast to the observation of
enhancement in visibility seen in the four-path experiment of Mei and
Weitz \cite{MW_1, MW_2}, it appears as though the three-path case shows no
such enhancement in visibility when one compares the case of complete path
indistinguishability ($\beta=1$) and that of complete path information
($\beta=0$) with the $\pi$ phase in the third path. However, note from
Fig. \ref{visibility_fig}(a) that if we look at an {\em intermediate} range
of values for $\beta$ (say, between $0.2$ and $0.5$), which corresponds
to {\em partial path information}, the $n=3$ case shows a regime in which
the visibility {\em enhances } with increasing path information (see
Fig. \ref{visibility_fig}(a)). 

In an earlier work, Bimonte and Musto constructed a three beam example to demonstrate
that one could have a peculiar situation where increasing path information
could lead to an increase in fringe visibility  \cite{bimonte1}.
However, their analysis fell short of demonstrating that  \cite{cohreview}.
The preceding analysis completes the task left unfinished in the analysis
of Bimonte and Musto. Clearly, this effect
is counterintuitive to the notion of complementarity as understood in
duality relations between conventional visibility and path information.
Unlike in the two-path case, in the three-path case, it is clear that
under conditions described above, the traditional visibility does not
capture the intuitive idea of wave-particle duality.
Since the preceding analysis shows that even in a three-beam interference
experiment, visibility can increase with increasing path information,
it also implies that the dulaity relation for three-slit 
interference, formulated by Siddiqui and Qureshi  \cite{asad}, will not
hold in such special scenarios.

Let us now analyze the observations reported in the four-path quantum
interference experiment of  Mei and Wietz  \cite{MW_1, MW_2}.  In this
experiment, a selective scattering of photons from just one interfering
path can lead to an {\em increase} in the fringe contrast if an additional phase of $\pi$ is included in the path. Although it is not possible to extract path information in this experiment,
it is obvious that the scattered photons do carry path information about the
particle. Our ancilla states can effectively mimic the role of photons.
From (\ref{intensity}), it can be shown that the intensity
of the quanton for $n=4$ is:
\begin{equation}\label{4_slits}
I= \vert \alpha \vert^{2} \left[1 + \frac{2-\beta}{2} \cos(\theta) + \frac{1-\beta}{2} \cos(2\theta)-\frac{\beta}{2} \cos(3\theta)\right].
\end{equation} 
From (\ref{4_slits}), the visibility as a function of the detector
overlap $\beta$ (which quantifies the path information) is shown in
Fig. \ref{visibility_fig}(b). Note that in the $n=4$ case, when there
is complete path information ($\beta=0$)  of the fourth path, the
visibility  {\em is more} than that of the case when all four paths are
indistinguishable ($\beta=1$). This is in agreement with the experimental
observations of Mei and Weitz. Also note that in the regime where partial
path information is there, one can see regions where there is a definite
enhancement of visibility with increasing path information. This trend
is more dramatic in the $n=4$ case compared to the $n=3$ case. It is
worth pointing out here that the inclusion of the $\pi$ phase is very
crucial to this effect since what would have been a perfect destructive
interference of the fourth path with the other three is now
being replaced by a more and more incoherent contribution
of this path to the interference pattern. It is easy to verify
from (\ref{I_nslit}) that the inclusion of a $\pi$ phase for
$n=2$, i.e., the  {\em two path} case will have no such effect.
Figs. \ref{visibility_fig}(c, d)  show similar plots for the cases $n=5$
and $n=6$, respectively, and one can see that an increase in the knowledge
of path information is accompanied by increase in the visibility of
the interference fringes. Upon further increasing the number of paths,
we see that the variation in the fringe visibility with $\beta$ (the
degree of path information of the $n^{th}$ path ) reduces and saturates
to a value close to $1$ as $n$ becomes very large.  This is akin to
a diffraction grating where larger number of slits leads to a sharper
fringe pattern or higher visibility. Clearly, as $n$ becomes very large,
the path information of just one path has a negligible effect on the
visibility as more and more paths coherently contribute to it. Thus one
can conclude that the counterintuitive effect seen in the experiment of
Mei and Weitz is significant only for a small number of interfering paths
and is most appreciable in the $n=4$ case (chosen for their experiment)
and becomes negligible as we increase $n$ beyond six paths.  Even so,
the enhancement in fringe contrast with increasing path information for
$n=3,4,5,6$ is in conflict with the spirit of Bohr's complementary and
points to the limitations of traditional visibility, (\ref{visibility}).

\subsection{Quantum Coherence}

\begin{figure}
\centerline{\resizebox{9cm}{!}{\includegraphics{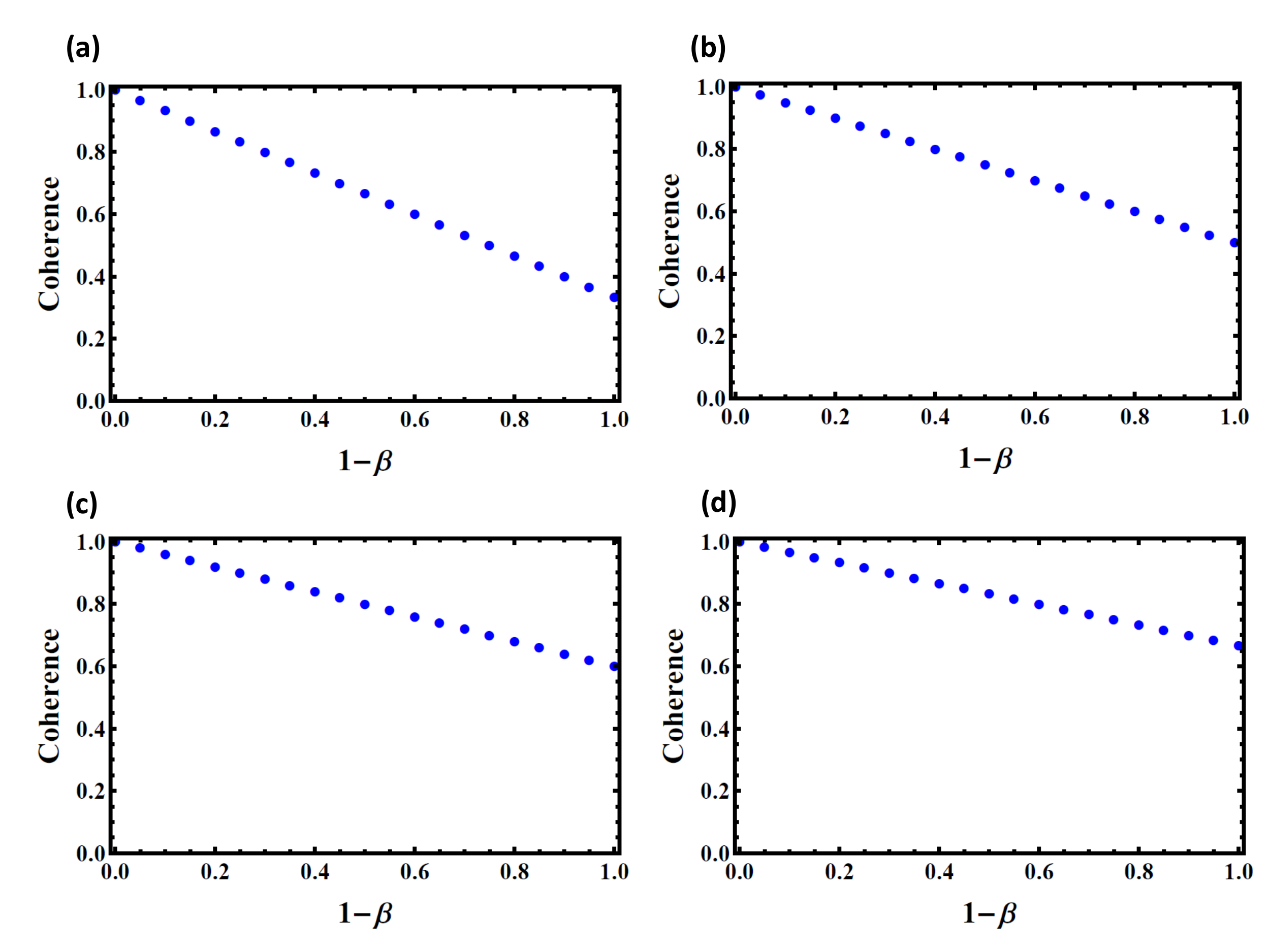}}}
\caption{Variation in coherence with respect to one-path knowledge ($1-\beta$)
for (a) 3 path (b) 4 path (c) 5 path  and (d) 6 path interference.}
\label{coherence_fig}
\end{figure}

Quantum entanglement and quantum coherence both  intimately capture the
intrinsic and uniquely quantum nature of systems.  A measure for  quantum
coherence, called the {\em $l_1$ norm}  recently defined in the framework
of quantum information theory by Baumgratz et al \cite{coherence} has
evoked a lot of interest and it has been argued that this {\em $l_1$ norm} is
a better quantifier of wave nature than traditional visibility. The
{\em $l_1$ norm} was generalized  to a normalized quantum coherence measure
by Bera {\em et al. } \cite{bera} to derive a  wave-particle duality relation
for arbitrary multi-path quantum interference phenomena.  It can be seen
that the coherence, (\ref{coherence}),  for the  state of the quanton,
(\ref{state}), is,
\begin{equation}
C(\rho_{red}) = \frac{2\beta}{n}+ \frac{n-2}{n}.
\label{coherence_n}
\end{equation}
We can see that for $n=2$, $C(\rho_{red})=\beta$ and for $n=3$ and
$n=4$, it is $\frac{2 \beta}{3} + \frac{1}{3}$ and $\frac{\beta+1}{2}$,
respectively. Note that the quantum coherence via $l_1$ norm is {\em
the same as the traditional visibility} for the two-path case. In Fig. 
\ref{coherence_fig} we show the variation of coherence with the degree of
one-path knowledge ($1-\beta$) for the cases of 3, 4, 5 and 6 paths.  Clearly,
unlike the visibility, the coherence  measured by the $l_1$ norm  {\em
decreases} monotonically  with increase in the path information, in the
spirit of Bohr's complementarity.  Note also that unlike  visibility,
the measure of coherence is unaffected by the presence or absence of
the $\pi$ phase in one path. (\ref{coherence_n}) also shows that when
the number of paths are large ($n\rightarrow \infty$), the degree of
which path information, $\beta$, is of little significance and coherence
approaches its maximum value of $1$.

For the illustration in the preceding sections,  we used a very simple
model to incorporate the effect of the $n$ detectors by assuming that
the overlaps of $n-1$ detectors is such that  $\langle \chi _{k}
\vert \chi _{j}\rangle=1$ and that these  $n-1$ detectors  states
all have an overlap with the $n$th detector state (which will provide
the which path information)  given by $ \langle \chi _{j} \vert \chi
_{n} \rangle = \beta$ where $j= 1,2,3 ..., n-1$.  In Mei and Weitz's
experiment \cite{MW_1, MW_2} done via multiple beam Ramsey interference
with Cesium atoms, while there are no explicit detector states entangled
with the quanton's state of four interfering paths, the degree of which
path information is presumably revealed by a scattering of photons from the fourth
path. In the experiment, the four interfering paths are represented
by the magnetic sublevels of the $F=3$  hyperfine component of the
Cesium electronic ground state.  Upon irradiating the atoms with resonant
optical beams
a coherent 
superposition of four magnetic ground state sublevels is created which
simulates the four path interference. The coherent superposition is probed
via atom interferometry techniques to obtain the fringe pattern. Between
the Ramsey pulses, one selected path is
coupled to the environment by a sequence of microwave pulses resulting
in path selective decoherence due to scattering of photons which can reveal
the which path  information. The initial motivation for Mei and Weitz's
experiment was to study decoherence in the multipath interference
scenario and they suggest from their counterintuitive observations that
in the case of multiple beam interference  the Michelson fringe contrast
(visibility) is not sufficient to quantify decoherence.
 
In the next section, we look at the multipath interference of Mei and
Weitz with path selective decoherence and $\pi$ phase by incorporating
the effect of a realistic environment and analyze its effect on visibility
and quantum coherence.

\section{Multislit interference with decoherence} \label{multislit_dech}

Let us now assume that as the particle (quanton) comes out of the n-slits
and travels to the screen, it is affected by weak interaction with
some kind of environment. We describe this environment as a reservoir
of non-interacting quantum oscillators, each of which interacts with
the particle. The Hamiltonian governing the particle can then be
represented as
\begin{equation}
H = \frac{p^2}{2m} + \sum_j {P_j^2\over 2M_j} 
    + {1\over 2}M_j \omega_j^2\left(X_j-{g_jx\over {M_j}\omega_j^2}\right)^2,
\end{equation} 
where $x,p$ are the position and momentum operators of the particle
{(quanton)},
$m$ is its mass, $X_j,P_j$ are position and momentum operators , and $M_j$
the mass
of the {jth} harmonic oscillator of frequency $\omega_j$ comprising the
environment and $g_j$ are the respective coupling strengths. Such a system
has been studied in great detail  \cite{legget,gsa,dekker,barchielli,dk,
av} and the solution for this Hamiltonian is governed by the master
equation given by
\begin{eqnarray}
\frac{\partial\rho(x,x',t)}{\partial t} &=& \left\{\frac{-\hbar}{2im}
\left( \frac{\partial^2}{\partial x^2}
- \frac{\partial^2}{\partial x'^2} \right)  \right.\nonumber\\
&& \left. -\gamma(x-x')\left( \frac{\partial}{\partial x} - \frac{\partial}{\partial x'} \right) 
- \frac{D}{4\hbar^2}(x-x')^2\right\}, \nonumber\\
\label{master}
\end{eqnarray}
where $\rho(x,x',t)$ is the reduced density matrix of the quanton
after tracing out the environmental degrees of freedom, $\gamma$ is the
Langevin friction coefficient, $D = 2m\gamma k_{B}T$ can be interpreted
as a diffusion coefficient, and $T$ is the temperature of the harmonic
oscillator
heat-bath \citep{legget,gsa,dekker}.

Let us assume that the state $\vert \psi_{k} \rangle$ which emerges from
the k’th slit, is a Gaussian wave-packet localized at the location of
the k’th slit, with a width equal to the width of the slit:
\begin{eqnarray}
\langle x|\psi_k\rangle &=& \tfrac{1}{(\pi/2)^{1/4}\sqrt{\epsilon}} e^{-(x-k\ell)^2/\epsilon^2},
\label{psik}
\end{eqnarray}
where $\ell$ is the distance between the centres of two neighboring slits,
and $\epsilon$ is their approximate width. Assuming that which-path
detectors present at each slit interact with the particle (quanton),
the combined wave-function of the particle and the detectors, as it
emerges from
 the n-slit{s},
is given by
\begin{eqnarray}
\langle x|\Psi\rangle &=& \tfrac{1}{(\pi/2)^{1/4}\sqrt{\epsilon}} 
\sum_{k=1}^n c_k e^{-(x-k\ell)^2/\epsilon^2}|\chi _{k} \rangle ,
\label{Psi}
\end{eqnarray}
The density matrix corresponding to this can be written as
\begin{eqnarray}
\rho_0(x,x') &=& \tfrac{1}{\sqrt{\pi/2}\epsilon}
\sum_{j,k} c_jc_k^* e^{\frac{-(x-j\ell)^2}{\epsilon^2}}
e^{\frac{-(x'-k\ell)^2}{\epsilon^2}}|\chi _{j} \rangle\langle \chi _{k} |.
\label{rho}
\end{eqnarray}
If one were to look only at the particle, it amounts to  tracing over
the states of the path detectors, giving us the reduced density matrix
for the quanton
\begin{eqnarray}
\rho(x,x',0) &=& \tfrac{1}{\sqrt{\pi/2}\epsilon}
\sum_{j,k} c_jc_k^* e^{\frac{-(x-j\ell)^2}{\epsilon^2}}
e^{\frac{-(x'-k\ell)^2}{\epsilon^2}}\langle \chi _{k} |\chi _{j} \rangle.
\label{rho0}
\end{eqnarray}
This is the density operator of the particle at time $t=0$ as it emerges
from the n-slit. Its decoherent dynamics will be governed by equation
(\ref{master}). Under the assumption that the effect of the environment
is so weak that
dissipative time-scales are much longer than the time the particle
takes to reach the screen ($t$)  after traveling a distance $L$, the
diagonal component of the density operator $\rho(x,x',t)$ representing
the probability density of the particle hitting the screen at a point $x$
can be given by
\begin{eqnarray}
\rho(x,x,t) &=& \frac{1}{\sqrt{\pi \alpha/2}}\left[\sum_{j=1}^n |c_j|^2
e^{-{2\epsilon^2(x-jl)^2/(\lambda L/\pi)^2}} \right.\nonumber\\
&& +\left. \sum_{j\ne k}
|c_j||c_k||\langle \chi _{k} |\chi _{j} \rangle| e^{-\epsilon^2\{(x-j\ell)^2 + (x-k\ell)^2\}\over(\lambda L/\pi)^2}{e^{\frac{-D(j-k)^2 \ell^2 t}{12 \hbar^2}}}  \right.\nonumber\\
&& \left. \cos\left\{\frac{2\pi\ell(k-j)
(x-\ell\frac{k+j}{2})}{\lambda L } + \theta_k-\theta_j\right\}\right],
\label{rho}
\end{eqnarray}
where $\alpha = \epsilon^2 + \frac{\hbar^2(1-e^{-2\gamma t})^2}{\epsilon^2m^2\gamma^2}
+ \frac{D[4\gamma t+4e^{-2\gamma t}-e^{-4\gamma
t}-3]}{8m^2\gamma^3}$. Staying within the Fraunhoffer limits, the width
of the slits ($\epsilon$) is much much small in comparison to the fringe
width ($ \lambda L / l$), so the narrow width of the Gaussians in (
\ref{rho0})  will become very large in  (\ref{rho}). So for all the
practical purposes, the values of all the Gaussians, at any point $x$
on the screen, is almost the same, and thus independent of $j, k$ and
the expression for $\rho(x,x,t)$ can be approximated as
\begin{eqnarray}
\rho(x,x,t) &\approx& \frac{e^{-{2\epsilon^2 x^2/(\lambda L/\pi)^2}}}{\sqrt{\pi \alpha/2}}\left[ 1 + \sum_{j\ne k}
|c_j || c_k||\langle \chi _{k} |\chi _{j} \rangle| {e^{\frac{-D(j-k)^2 \ell^2 t}{12 \hbar^2}}}  \right.\nonumber\\
&& \left. \cos\left\{\frac{2\pi\ell(k-j)x}{\lambda L } + \theta_k-\theta_j\right\}\right].
\label{rhoa}
\end{eqnarray}

(\ref{rhoa}) is a recently reported result by Venugopalan {\em et al. }
\cite{venu_2019} for an $n$ slit interference in the presence of
environment induced decoherence. Note that in the limit of coupling
with the
environment going to zero, which  {is the regime when decoherence is
completely absent}, i.e., when $\gamma\to 0,~D\to 0$,  (\ref{rhoa})
reduces to eqn. (16) of ref. \cite{tania} which is the known result for
a multi path interference.
(\ref{rhoa}) shows that the  decohering n-slit interference pattern
is also built up from all possible {\em two-slit inteference terms}
which is also the case when there is no coupling to the environment
\cite{tania}. The environmental coupling has had the effect of modifying
these pair-wise contributions and this effect of decoherence is neatly
condensed into
the exponential factor {$e^{-D(j-k)^2 \ell^2 t/12 \hbar^2}$}  in
(\ref{rhoa}) which multiplies
the cosine term giving rise to the interference. Notice that the exponential
decay {term} in (\ref{rhoa})  whose argument 
is {$-D(j-k)^2 \ell^2 t/12 \hbar^2$} 
cannot be pulled out of the summation,
{as} it depends on j, k. It may also be noted  that
for each pair of slits, there will be a characteristic decoherence time, $\tau_{d}^{(jk)} =
12\hbar^2/D(j-k)^2 \ell^2$, and for n-slits
there will be a total of $n(n-1)/2$ time scales which will collectively contribute in the summation leading to the degradation of the {$n$-slit } pattern. For the simplest case of two slits, there will be only one time scale, $\tau_{d} = 12\hbar^2/D \ell^2$.

Now let us see how we can apply the general result (\ref{rhoa}) to the
situation explored in  the experiment of Mei and Weitz. As discussed in
detail in the previous sections,  in the $n$ path interference, there is
an additional $\pi$ phase added to the $n^{th}$  path and  {there is also
the possibility of which path information from decoherence
via scattering of photons for this  path}.  In the present model, decoherence will be
instrumental in revealing the which path information, and we assume that
the effect of the environment {\em applies only to the $n^{th}$ path}.
Note that this means we set all detector overlaps to $1$, making them
irrelevant for now. With these conditions taken into account, the
probability density of the particle hitting the screen at a point $x$
can be written as
\begin{eqnarray}
\rho(x,x,t) &\approx & \frac{e^{-{2\epsilon^2 x^2/(\lambda L/\pi)^2}}}{\sqrt{\pi \alpha/2}} \left[1   + \sum_{j\ne k} ^{n-1}
|c_j || c_k| \right.\nonumber\\
&& \left. \cos\left\{\frac{2\pi\ell(k-j)x}{\lambda L } + \theta_k-\theta_j\right\} \right.\nonumber\\
&& - \left. 2 \sum_{j} ^{n-1}
\vert c_j \vert \vert c_n \vert \hspace{1mm}  \hspace{1mm} {e^{\frac{-D(n-j)^2 \ell^2 t}{12 \hbar^2}}}  \right.\nonumber\\
&&  \left. \cos\left\{\frac{2\pi\ell(n-j)x}{\lambda L } + \theta_n-\theta_j\right\}\right].
\label{rhob}
\end{eqnarray}
({\ref{rhob}) captures the multipath interference pattern as the position
probability distribution on the screen when the $n^{th}$ path is impacted
by environmental influence and has an additional $\pi$ phase. Clearly,
with the passage of time, one can see progressive decoherence happening
from the contribution of the  last term  which eventually kills the
interference between the $n^{th}$ path and all the others, increasingly
revealing which path information.  When $t \rightarrow \infty$, there
is complete path information of the $n^{th}$ path. With $n=4$, one can
explore (\ref{rhob}) to explain the results of Mei and Weitz which we
will do in the next section.

Next, we ask if it is possible to extract the amount of quantum coherence
as defined in (\ref{coherence}) from the interference  (\ref{rhob})
and have a way of measuring it in a real experiment. It has been shown
by Paul and Qureshi \cite{tania} that it is possible to measure quantum
coherence in a real experiment for a multislit interference with which
way detectors whose path-distinguishability is tunable
\cite{tania} 
between two modes: (a) one that makes all paths completely
{\em indistinguishable} and (b) one that makes  all  paths fully
{\em distinguishable}. If the two cases, (a) and (b), are denoted by
$\parallel$ and $\perp$, respectively, Paul and Qureshi provide the
following protocol: First,
the intensity at a primary maximum $I_{max}^{\parallel}$ is measured when
the {$n $} paths are indistinguishable, i.e., $|\chi_i\rangle$s are all
identical
and parallel.  Next, the path-detector is switched to the mode (b)
where all the $n$ paths
are fully distinguishable, and the intensity $I_{max}^{\perp}$ is
measured at
the same location on the screen as before.  Coherence of the incoming
{quanton} can then be measured as \cite{tania}
\begin{eqnarray}
\mathcal{C}_{expt} = \frac{1}{n-1}\frac{I_{max}^{\parallel} - I_{max}^{\perp}}{I_{max}^{\perp}} .
\label{Cexpt}
\end{eqnarray}

While this protocol works for most cases of  $n$ path
interference, in an experiment like that of Mei
and Weitz, notice that there is a deliberate arrangement of the $\pi$
phase in one particular path  such
that the knowledge of path information for this path,   instead of
degrading the interference pattern (as normally expected), leads to
a sharper contrast, as has been elaborated in the previous sections.
In such  situations, the protocol of Paul and Qureshi fails to provide
an experimental measure of coherence. However,
we have seen from  discussions in the previous sections  that coherence
(\ref{coherence})  does indeed decrease with increasing path information,
even while the interference pattern becomes sharper.  How then, can
we measure the coherence? The way around this is a recent result by
Qureshi \cite{coh_exp} where a modified and unconventional protocol
is proposed which allows for the measurement of quantum coherence in
situations such as the experiment of Mei and Weitz where there could
be constraints on phases. Coherence can be measured in such cases by
opening only one pair of paths at a time (while blocking the rest) ,
and measuring conventional
visibility for each pair at a time and finally
taking an average over all  $n(n-1)/2$  pairs of paths. Essentially, this
casts quantum coherence as a sum of visibilities of interference from
individual pairs of path. We can show that  coherence (\ref{coherence})
of the pattern (\ref{rhob}) can then be measured in terms of pair wise
visibility using this protocol \cite{coh_exp} as:
\begin{eqnarray}
C(\rho(x,x,t)) &=& \frac{2}{n(n-1)}\sum_{j\ne k} ^{n-1} \frac{2|c_j || c_k|}{|c_j |^{2}+ |c_k|^{2}} \nonumber\\
 &+& \frac{2}{n(n-1)} \sum_{j} ^{n-1} \frac{2|c_j || c_n|}{|c_j |^{2}+ |c_n|^{2}} {e^{\frac{-D(n-j)^2 \ell^2 t}{12 \hbar^2}}}.
\label{coha}
\end{eqnarray}
We recall, once again, that the two-path scenario remains the exceptional
case where quantum coherence is the same as traditional visibility. In
the next section we discuss and analyze the results (\ref{rhob})
and (\ref{coha}).

\section{Results and discussion} \label{results}

In the previous sections, we have analyzed the multi slit interference
set up with which way detectors for situations where there is a
possibility of which way path detection for one particular path
either through the  detector or by means of path selective decoherence.
Additionally,  in tune with the experiment of Mei and Weitz,  the selected
path has an extra phase of $\pi$. In the preceding section, incorporating
a model for the environment, we have obtained an expression for the
intensity distribution of the quanton, (\ref{rhob}), and the quantum
coherence (\ref{coha}) in terms of the parameters of the environment. Let
us now try to analyze how the visibility and coherence of the system
evolves with time for the case of an initial maximally coherent state,
(\ref{state}), i.e., when $c_{i}= \frac{1}{\sqrt{n}}$. The intensity
distribution on the screen of the quanton is then
\begin{eqnarray}
\rho(x,x,t) &\approx& \frac{e^{-{2\epsilon^2 x^2/(\lambda L/\pi)^2}}}{\sqrt{\pi \alpha/2}}\left[ 1 +  \frac{1}{n}\sum_{j\ne k} ^{n-1}
 \cos\left\{\frac{2\pi\ell(k-j)x}{\lambda L } + \theta_k-\theta_j\right\} \right.\nonumber\\
&& - \left. \frac{2}{n} \sum_{j} ^{n-1} {e^{\frac{-D(n-j)^2 \ell^2 t}{12 \hbar^2}}} \cos\left\{\frac{2\pi\ell(n-j)x}{\lambda L } + \theta_n-\theta_j\right\}\right],
\label{rhof}
\end{eqnarray}
and the quantum coherence is
\begin{eqnarray}
C(\rho(x,x,t)) = \frac{n-2}{n} + \frac{2}{n(n-1)}\sum_{j} ^{n-1} {e^{\frac{-D(n-j)^2 \ell^2 t}{12 \hbar^2}}}.
\label{cohb}
\end{eqnarray}
In  Fig. \ref{neon_4_slits} we plot the intensity pattern of the quanton
with the passage of time,  for the $n=4$  case. Clearly, the pattern
shows an enhancement of fringe contrast with increasing decoherence, as
observed in the experiment of Mei and Weitz.  The variation of visibility
and coherence with time for  $n=3, 4, 5$ and $6 $ are plotted in Fig. \ref{neon_vis_coh}.
\begin{figure}
\centerline{\resizebox{9cm}{!}{\includegraphics{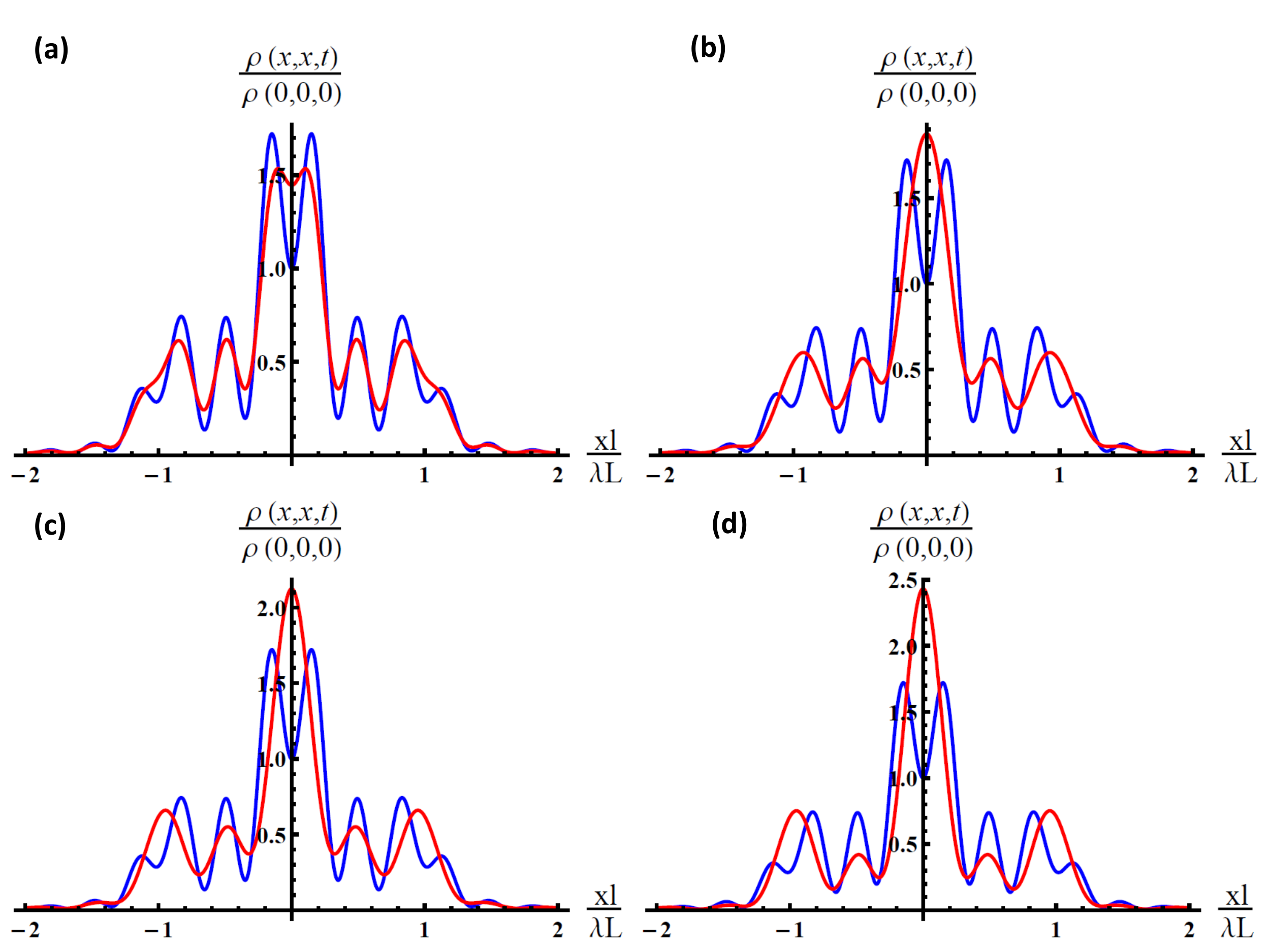}}}
\caption{Plots of $\rho(x,x,t)/\rho(0,0,0)$ at the screen, for 4 path
interference, showing how decoherence progressively affects interference.
The following parameters used are adapted from the interference experiment
on ultracold Neon atoms \cite{neon}:
$m=3.349\times 10^{-26}~\text{Kg}, T = 2.5~\text{mK}, \lambda=0.018~\mu\text{m},
\ell=6~\mu\text{m}, L=37~\text{mm}$.
The blue plots are $t/\tau_d=0$ while the red plots are for (a)
$t/\tau_d=1/12$, (b) $t/\tau_d=1/4$, (c) $t/\tau_d=1/2$ and (c)
$t/\tau_d=2$ where $\tau_{d} = 12\hbar^2/D \ell^2$. We can observe that
with increase in time the there is an enhancement of the visibility. }
\label{neon_4_slits}
\end{figure}

\begin{figure}
\centerline{\resizebox{9cm}{!}{\includegraphics{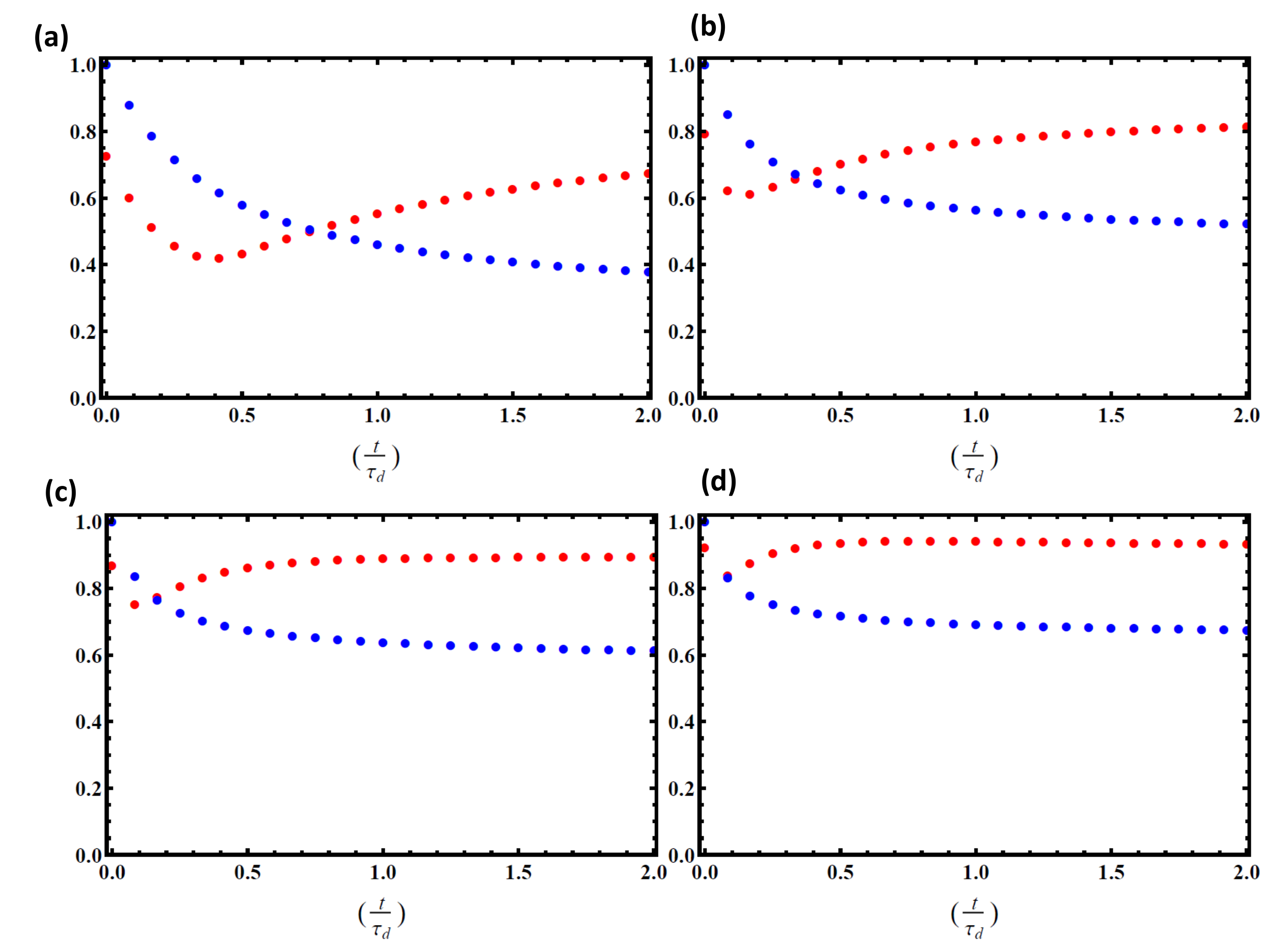}}}
\caption{Variation of visibility (in red) and  coherence (in blue) with respect
to scaled time $t/\tau_d$ for (a) 3 path (b) 4 path (c) 5 path and (d) 6 path interference.}
\label{neon_vis_coh}
\end{figure}

Let us now examine the initial visibility and quantum coherence of
the quanton before decoherence starts in the $n^{th}$ path. Note that
since all the which way detectors are  assumed to be identical, and the
quanton is initially in a pure state,  at $t=0$  the  initial quantum
coherence of the quanton for all cases  has the maximum value of $1$.
However,  the visibility at $t=0$ does not have its maximum value of $1$. This is because the  presence of  an additional  $\pi$ phase in one of the paths affects the  visibility and keeps it at a value that is  less than $1$ at $t=0$. Thus, unlike coherence, even all identical which way
detectors and  a pure maximally coherent state does not ensure maximum
visibility. Coherence is unaffected by the presence of the $\pi$ phase.
With the passage of time, there is path selective  decoherence in one
of the paths  and this affects the overall interference pattern and is
also the mechanism for progressively revealing the path information. At
$t=0$ there is no decoherence and at  $t\gg \frac{12 \hbar^2}{Dl^2}$,
i.e., $t\gg \tau_d$ we have a situation where there is full decoherence. For the $n$
path case, the coherence (\ref{cohb}) of the quanton decays and saturates
to the value of $\frac{n-2}{n}$ for  large  times when the decoherence
has played its role to the fullest. Note that this was also seen in
(\ref{coherence_n}).

Let us now look at the  specific case of the $3$ path interference
(see Fig. \ref{neon_vis_coh}(a)). We can see that with the passage of
time,  while the coherence of the quanton decays from a maximum value
of $1$ to minimum  value of $1/3$, the visibility shows a rather unusual
variation. For short times, the plot shows a decrease in visibility but
as time progresses there is an increase in the visibility. Thus, even
though there is progressive decoherence over time, and hence increasing
path information, there is a definite observable {\em increase } in the
visibility  which clearly goes against the spirit of what is understood
conventionally as  Bohr's complementary principle. Note that this was
also seen in our analysis in Section \ref{multislit} (see Fig. \ref{visibility_fig}(a)). Similarly,  for  $4$
paths, one can see from Fig. \ref{neon_vis_coh}(b) that with the passage
of time while the coherence of the quanton decays from a maximum value of $1$
to a minimum  value of $1/2$, the visibilty of the interference fringes
{\em increases}, as seen in the experiment of Mei and Weitz. Also, for
this case, it is clear from our analysis that the visibility for full path
information (large $t$)  is {\em greater} than in the case of no path
information ($t=0$). A similar trend for  visibilty and coherence can
be seen for $5 $ and $6$ path  cases (see Fig. \ref{neon_vis_coh}(c,d)),
though they are less dramatic as compared to the $n=4$ case. Note that
with the increase in the number of paths, the effect of decoherence in
one of the paths on the visibility and coherence decreases and both
saturate to a value practically close to the maximum value $1$. This
can be understood from the fact that with large number of available
paths the effect of decoherence in just one of the paths will have
a negligible impact on the overall interference pattern, and a large
number of interfering paths contribute to the enhancement of visibility
and coherence.

\section{Conclusions} \label{conclusions}
To summarize, we have  analyzed multi-path quantum interference where
which path information for one particular path is revealed by
which way detectors with an additional $\pi$ phase in the selected path. 
We have also analyzed  multi-path quantum interference where a 
path selective  decoherence plus an additional $\pi$ phase
in the selected path, is introduced, as in the experiment of Mei and Weitz.
Through this analysis we explain the
observations of an experiment by Mei and Weitz where path selective
decoherence can lead to, not only a decrease, but an {\em increase}
of the fringe contrast. The effect of the environment is modeled via
a coupling to a bath of harmonic
oscillators. When the effect of the environment (decoherence) is
introduced in one of the four paths, an  enhancement in fringe contrast is
seen under the conditions of \cite{MW_1, MW_2}. We show that  a similar
effect can be seen in the general case of multi-path interference with
path selective decoherence and illustrate it additionally for three, five
and six slits.
Our results suggest  that traditional fringe visibility, while capturing well
the intuitive idea of complementarity in the two-path case,  fails in
multi path situations such as those seen in the experiment of Mei and
Weitz, raising serious doubts on its role as an indicator of coherence
and wave nature. We also probe the effect seen in \cite{MW_1, MW_2}
using the recently introduced $l_1$ norm of {\em quantum coherence} which
can be experimentally measured by a recently reported protocol
\cite{coh_exp}.  We explain the enhancement of fringe visibility and show
that even in situations where visibility could increase with increasing
decoherence, or increasing which path information, coherence always
decreases and preserves its role as a quantifier of wave nature. 
The observation of increased  fringe contrast in Mei and Weitz's experiments is for
a four path interference. A similar effect can be seen in certain
regimes for three path interference, and also for five and six paths.
In  \cite{MW_1, MW_2} it is suggested that the four path experiment could
be extended towards an increased number of interfering paths or quantum
systems of larger size. Our observation here is that the visibility
enhancement effect is interesting only for the cases of a few paths
and increasing the number of paths  beyond these numbers tones down
this dramatic effect. However, interference with few paths with larger
quantum systems would certainly be interesting. In \cite{MW_1, MW_2} it is
suggested that traditional visibility fails as a signature for quantifying
decoherence. Indeed, the enhancement of fringe contrast in the presence
of environmental decoherence highlights the limitations of traditional visibility as a good measure for wave nature in complementarity and makes it an unlikely candidate for quantifying
decoherence. Our results affirm the fact that the amount of decoherence
can only be estimated by a measure of the residual coherence and cannot
be  inferred by changes in fringe contrast. This normalized $l_1$ norm
of coherence can now be considered as a true and legitimate measure of
the wave aspect of the quanton in a multi path experiment. Further,
this quantum coherence can be measured experimentally in an $n$ path
interference experiment by taking the average of {\em traditional
visibility} for all available pairs of two path interference by blocking
the rest of the $n-2$ paths \cite{coh_exp}. Our results and analysis
could lead to better insight and understanding  of wave-particle duality
and complementarity, visibility and quantum coherence in multi path
interference and will be of significance in studies  that seek to exploit
quantum superpositions and quantum coherence for quantum information
applications.

\section*{Acknowledgment}
Sandeep Mishra thanks Guru Gobind Singh Indraprastha University, Delhi for the
Short Term Research Fellowship.

\end{document}